\documentclass[preprint,aps,prl,showpacs]{revtex4}
\usepackage{times}
\usepackage{amstext,amsfonts,amsbsy,amssymb,bbm,latexsym}
\usepackage{graphicx}% Include figure files
\def\Tr{\mathop{\rm Tr}\nolimits}

\newcommand{\pref}[1]{(\ref{#1})}

\newcommand{\iden}{\mathbbm{1}}

\newcommand\ie {{\it i.e.}, }

\newcommand{\nn}{\nonumber\\}

\newcommand{\bs}{\boldsymbol}

\newcommand\ra{\rightarrow }

\newcommand\half{\frac 1 2 }

\renewcommand{\v}[1]{{\bf #1}}

\newcommand{\ket}[1]{|#1\rangle}
\newcommand{\bra}[1]{\langle #1 |}
\newcommand{\braket}[2]{\langle #1|#2 \rangle}
\newcommand{\ketbra}[2]{|#1\rangle\langle#2|}

\newcommand{\cB}{ {\cal B} }

\newcommand{\cD}{ {\cal D} }
\newcommand{\cE}{ {\cal E} }

\newcommand{\cP}{ {\cal P} }

\newcommand{\cS}{ {\cal S} }

\newcommand {\be}[1]{\begin{eqnarray} \mbox{$\label{#1}$}  }
\newcommand{\ee}{\end{eqnarray}}

\begin{document}

\title{Geometrical aspects of entanglement}
\author{Jon Magne Leinaas$^{a}$, Jan Myrheim$^{c}$ and Eirik Ovrum$^{a,b}$}
\affiliation{${(a)}$ Department of Physics,
University of Oslo, P.O.\ Box 1048~Blindern, 0316~Oslo, Norway}
\affiliation{${(b)}$ Centre of Mathematics for Applications,
P.O.\ Box 1053~Blindern, 0316 Oslo, Norway}
\affiliation{${(c)}$ Department of Physics,
The Norwegian University of Science and Technology,
7491~Trondheim, Norway}

%\date{\today}
\date{May 8, 2006}

\begin{abstract}
  We study geometrical aspects of entanglement, with the
  Hilbert--Schmidt norm defining the metric on the set of density
  matrices.  We focus first on the simplest case of two two-level
  systems and show that a ``relativistic'' formulation leads to a
  complete analysis of the question of separability.  Our approach is
  based on Schmidt decomposition of density matrices for a composite
  system and non-unitary transformations to a standard form.  The
  positivity of the density matrices is crucial for the method to
  work.  A similar approach works to some extent in higher dimensions,
  but is a less powerful tool.  We further present a numerical method
  for examining separability, and illustrate the method by a numerical
  study of bound entanglement in a composite system of two three-level
  systems.
\end{abstract}

\maketitle

%%%%%%%%%%
\section{Introduction}
%%%%%%%%%%

Entanglement is considered to be one of the main signatures of quantum
mechanics, and in recent years the study of different aspects of
entanglement has received much attention. One approach has been to
study the formal, geometrical characterization of entangled states as
opposed to non-entangled or separable states
\cite{Kus01,Verstraete02,Pittenger03}. In such a geometrical approach
the Hilbert--Schmidt metric defines in many ways the natural metric on
the space of of physical states. This metric follows quite naturally
from the Hilbert space norm, when the quantum description is extended
from pure to mixed states, and it is a Euclidean metric on the set of
density matrices.  For a composite quantum system the separable states
form a convex subset of the full convex set of density matrices, and
one of the aims of the geometrical approach is to give a complete
specification of this set and thereby of the non-separable or
entangled states.

The purpose of this paper is to examine some questions related to the
geometrical description of entanglement. We focus primarily on the
simplest composite systems consisting of two two-level systems
($2\times 2$ system) or two three-level systems ($3\times 3$ system),
but examine also some questions relevant for higher dimensions.  In
the case of two two-level systems the separable states can be fully
identified by use of the Peres criterion \cite{Peres96}. This
criterion states that every separable density matrix is mapped into a
positive semidefinite matrix by partial transposition, \ie by a
transposition relative to one of the subsystems.  Since also
hermiticity and trace normalization is preserved under this operation,
the partial transpose of a separable density matrix is a new density
matrix.

A non-separable density matrix, on the other hand, may or may not
satisfy Peres' condition.  This means that, in general, Peres'
condition is necessary but not sufficient for separability. However,
for the special case of a $2\times 2$ system as well as for a $2\times
3$ system the connection is stronger, and the Peres condition is both
necessary and sufficient for a density matrix to be separable
\cite{Horodecki96}. To identify the separable density matrices, and
thereby the entangled states, is therefore in these cases relatively
simple. In higher dimensions, in particular for the $3\times 3$
system, that is not the case.  Peres' condition is there not
sufficient for separability, and the convex subset consisting of all
the density matrices that satisfy this condition is larger than the
set of separable matrices.  This is part of the reason that the
identification of the set of separable states in higher dimensions is
a hard problem \cite{Gurvits03}.

In the present paper we first give, in Sect.~2, a brief introduction
to the geometry of density matrices and separable states. As a next
step, in Sect.~3 we focus on the geometry of the two-level system and
discuss the natural extension of the Euclidean three dimensional
Hilbert--Schmidt metric to a four dimensional indefinite Lorentz
metric. This indefinite metric is useful for the discussion of
entanglement in the $2\times 2$ system, where Lorentz transformations
can be used to transform any density matrix to a diagonal standard
form in a way which preserves separability and the Peres condition
(Sect.~4).  By using this standard form it is straightforward to
demonstrate the known fact that any matrix that satisfies the Peres
condition is also separable.

The transformation to the diagonal standard form is based on an
extension of the Schmidt decomposition to the matrix space with
indefinite metric. For general matrices such a decomposition cannot be
done, but for density matrices it is possible due to the positivity
condition. We show in Sect.~5 that the Schmidt decomposition in this
form can be performed not only for the $2\times 2$ system, but for
bipartite systems of arbitrary dimensions. However, only for the
$2\times 2$ system the decomposition can be used to bring the matrices
to a diagonal form where separability can be easily demonstrated. In
higher dimensions the Schmidt decomposition gives only a partial
simplification. This indicates that to study separability in higher
dimensions one eventually has to rely on the use of numerical methods
\cite{Ioannou04,Bruss05,Horodecki00,Doherty02,Badziag05}.

In Sect.~6 we discuss a new numerical method to determine 
separability
\cite{Dahl06}, and use the method to study entanglement 
in the $3\times 3$ 
system. The method is based on a
numerical estimation of the Hilbert--Schmidt distance from any chosen
density matrix to the closest separable one. We focus particularly on
states that are non-separable but still satisfy the Peres condition.
These are states that one usually associates with bound entanglement.
A particular example of such states is the one-parameter set discussed
by P.~Horodecki \cite{Horodecki97} and extended by Bruss and Peres
\cite{Bruss00}. We study numerically the density matrices in the
neighbourhood of one particular Horodecki state and provide a map of
the states in a two dimensional subspace.

%%%%%%%%%%
\section{The geometry of density matrices}
%%%%%%%%%%
\subsection{The convex set of density matrices}
%%%%%%%%%%

A density matrix $\rho$ of a quantum system has the following properties,
\be{densmat}
\rho^{\dag}&=&\rho\quad\text{hermiticity}\nn
\rho&>&0\quad \text{positivity}\nn
\text{Tr}\;\rho&=&1\quad \text{normalization}
\ee
The matrices that satisfy these conditions form a convex set, and can
be written in the form
\be{conmat}
\rho=\sum_k p_k\, \ket{\psi_k}\bra{\psi_k}
\ee
with $\ket{\psi_k}$ as Hilbert space unit vectors and
\be{prob}
p_k \geq 0,\quad \sum_k p_k=1
\ee
The coefficient $p_k$, for given $k$, can be interpreted as the 
probability that the
quantum system is in the pure state $\ket{\psi_k}$.  This
interpretation depends, however, on the representation~\pref{conmat},
which is by no means unique.  In particular, the vectors
$|\psi_k\rangle$ may be chosen to be orthonormal, then they are
eigenvectors of $\rho$ with eigenvalues $p_k$, and
Eq.~(\ref{conmat}) is called the {\em spectral representation} of
$\rho$.

The pure states, represented by the one dimensional projections
$\ket{\psi}\bra{\psi}$, are the extremal points of the convex set of
density matrices.  That is, they generate all other density matrices, 
corresponding to
mixed states, by convex combinations of the form \pref{conmat},
but cannot themselves be expressed as non-trivial convex combinations of other
density matrices.

The hermitian matrices form a real vector space with a natural scalar
product, $\text{Tr}(AB)$, which is bilinear in the two matrices $A$
and $B$ and is positive definite. From this scalar product a Euclidean
metric is derived which is called the Hilbert--Schmidt (or Frobenius) 
metric on the
matrix space,
\be{dist}
ds^2={1\over 2}\,\Tr((d\rho)^2)
\ee
The scalar product between pure state density matrices
$\rho_1=\ket{\psi}\bra{\psi}$ and $\rho_2=\ket{\phi}\bra{\phi}$ is
\be{scalprod2}
\Tr(\rho_1\rho_2)=|\braket{\psi}{\phi}|^2
\ee
For an infinitesimal displacement $\ket{d\psi}$ on the unit sphere in
Hilbert space the displacement in the matrix space is
\be{drhodpsi}
d\rho
=\ketbra{d\psi}{\psi}+\ketbra{\psi}{d\psi}
\ee
and the Hilbert--Schmidt metric is
\be{dist2}
ds^2=\braket{d\psi}{d\psi}-|\braket{\psi}{d\psi}|^2
\ee
where we have used that
$\Tr(d\rho)=\braket{\psi}{d\psi}+\braket{d\psi}{\psi}=0$. This may be
interpreted as a metric on the complex projective space, called the
Fubini--Study metric.  It is derived from the Hilbert space metric and
is a natural measure of distance between pure quantum states, in fact,
$ds$ is the infinitesimal angle between rays (one dimensional
subspaces) in Hilbert space.  Since the Hilbert--Schmidt metric on
density matrices is a direct extension of the Hilbert space metric, it
is a natural metric for all, both pure and mixed, quantum states.

A complete set of basis vectors $\{J_a\}$, in the space of hermitian
matrices, can be chosen to satisfy the normalization condition
\be{basis}
\Tr(J_aJ_b)=\delta_{ab}
\ee
For $n\times n$ matrices the dimension of the matrix space, and the
number of basis vectors, is $n^2$.  One basis vector, $J_0$, can be
chosen to be proportional to the identity $\iden$, then the other
basis vectors are traceless matrices. A general density matrix can be
expanded in the given basis as
\be{densmatexp}
\rho=\sum_a \xi_a J_a
\ee
where the coefficients $\xi_a$ are real, and the trace normalization
of $\rho$ fixes the value of $\xi_0$.  With the chosen normalization
of $J_0$ we have $\xi_0=1/\sqrt{n}$.

\begin{figure}[h]
\begin{center}
\includegraphics[width=7cm]{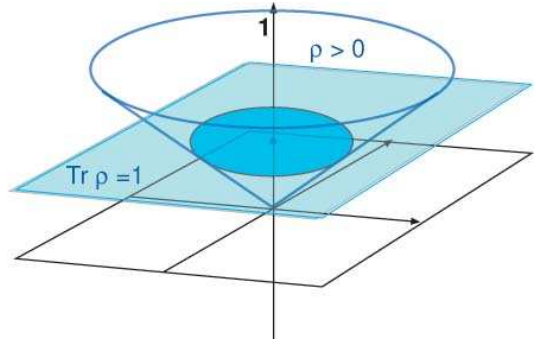}
\end{center}
\caption{{\small A schematic representation of the set of density
    matrices in the vector space of hermitian matrices. The positive
    matrices form a cone about the axis defined by the unit matrix,
    and the normalization condition restricts the density matrices to
    a convex subset, here represented by the shaded circle.} }
\label{geometry}
\end{figure}

Due to the normalization, the density matrices are restricted to a
hyperplane of dimension $n^2-1$, shifted in the direction of $J_0$
relative to a plane through the origin. The set of density matrices is
further restricted by the positivity condition, so it forms a closed,
convex set centered around the point $\rho_0=J_0/\sqrt{n}$. This point
corresponds to the {\em maximally mixed state}, which has the same
probability $\bra{\psi}\rho_0\ket{\psi}=1/n$ for any pure state
$\ket{\psi}$. The geometry is schematically shown in
Fig.~\ref{geometry}, where the set of density matrices is pictured as
the interior of a circle.  One should note that the normalization
condition in a sense is trivial and can always be corrected for by a
simple scale factor.  In the discussion to follow we will find it
sometimes convenient to give up this constraint. The quantum states
can then be viewed as rays in the matrix space, and the positivity
condition restricts these to a convex sector (the cone in
Fig.~\ref{geometry}).

%%%%%%%%%%%%
%%%%%%%%%%%%

%%%%%%%%%%
\subsection{Unitary transformations}
%%%%%%%%%%

The hermitian matrices $J_a$ can be viewed as generators of unitary
transformations,
\be{unitar}
U=\exp\!\left(i \sum_a \zeta_a J_a\right)
\ee
with real coefficients $\zeta_a$, which act on the
density matrices in the following way,
\be{adjoint}
\rho\ra\tilde{\rho}=U\rho U^{\dag}
\ee
If $\rho$ is represented as in Eq.~\pref{conmat}, then we see
that
\be{conmatU}
\tilde{\rho}=\sum_k p_k\,
\ket{\tilde{\psi}_k}\bra{\tilde{\psi}_k}
\ee
where $\ket{\tilde{\psi}_k}=U\ket{\psi_k}$.  Thus, the matrix
transformation $\rho\ra U\rho U^{\dag}$ is induced by the vector
transformation $\ket{\psi}\ra U\ket{\psi}$.  An immediate consequence
of Eq.~\pref{conmatU} is that the transformed density matrix $U\rho
U^{\dag}$ is positive.

Such unitary transformations respect both the trace and positivity
conditions and therefore leave the set of density matrices
invariant. Also the von Neumann entropy
\be{entro}
S=-\Tr(\rho \ln\rho)
\ee
is unchanged, which means that the degree of mixing, as measured by
the entropy, does not change.

Since the dimension of the set of density matrices grows quadratically
with the Hilbert space dimension $n$ the geometry rapidly gets
difficult to visualize as $n$ increases.  However, the high degree of symmetry
under unitary transformations simplifies the picture.  The unitary
transformations define an $SU(n)$ subgroup of the rotations in the $n^2-1$
dimensional matrix space, and all density matrices can be obtained
from the diagonal ones by these transformations. In this sense the
geometry of the set of density matrices is determined by the geometry
of the set of {\em diagonal} density matrices. The diagonal matrices
form a convex set with a maximal set of $n$ commuting pure states as
extremal points. Geometrically, this set is a regular hyperpyramid, a
{\em simplex}, of dimension $n-1$ with the pure states as corners. The
geometrical object corresponding to the full set of density matrices
is generated from this by the $SU(n)$ transformations.

\begin{figure}[h]
\begin{center}
\includegraphics[width=10cm]{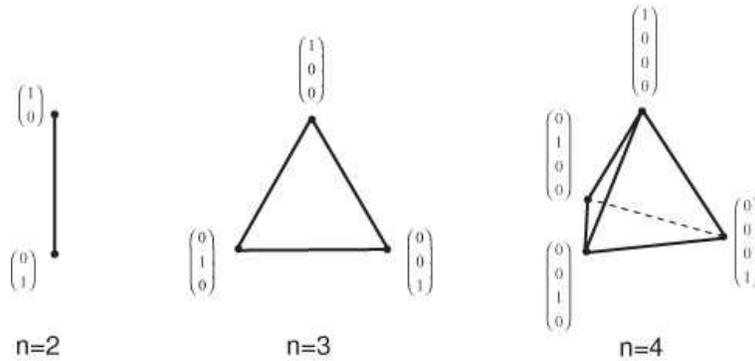}
\end{center}
\caption{{\small Geometric representation of the diagonal density
     matrices for the three cases of Hilbert space dimension 1, 2 and
     3. For general dimension $n$ they define a hyperpyramid of
     dimension $n-1$.} }
\label{pyramid}
\end{figure}

In Fig.~\ref{pyramid} the set of diagonal density matrices is
illustrated for $n=2,3$ and $4$, where in the first case the
hyperpyramid has collapsed to a line segment, for $n=3$ it is an
equilateral triangle and for $n=4$ it is a tetrahedron. For $n=2$, the
$SU(2)$ transformations generate from the line segment the three
dimensional {\em Bloch sphere} of density matrices. This case is
special in the sense that the pure states form the complete surface of
the set of density matrices. This does not happen in higher
dimensions. In fact, the dimension of the set of pure states is
$2n-2$, the dimension of $SU(n)/(U(1)\times SU(n-1))$, because one
given pure state has a $U(1)\times SU(n-1)$ invariance group.  This
dimension grows linearly with $n$, while the dimension of the surface,
$n^2-2$, grows quadratically.

%%%%%%%%%%%%
%%%%%%%%%%%%

The faces of the hyperpyramid of dimension $n-1$ are hyperpyramids of
dimension $n-2$, corresponding to density matrices of the subspace
orthogonal to the pure state of the opposite corner. Similarly, the
hyperpyramid of dimension $n-2$ is bounded by hyperpyramids of
dimension $n-3$, etc. This hierarchy is present also in the full set
of density matrices, generated from the diagonal ones by $SU(n)$
transformations. Thus, to each extremal point (pure state) the
boundary surface opposite to it is a flat face corresponding to the
set of density matrices of one lower dimension. In this way the
boundary surface of the set of density matrices contains a hierarchy
of sets of density matrices of lower dimensions.

The boundary of the set of density matrices is characterized by at
least one of the eigenvalues of the density matrices being zero, since
outside the boundary the positivity condition is broken. This means
that at the boundary the density matrices satisfy the condition
$\det\rho=0$, which is an algebraic equation for the coordinates of
the boundary points. When $n$ is not too large the equation can be
solved numerically. This has been done in Fig.~\ref{Bound} where a two
dimensional section of the set of density matrices is shown.  One
should note that there will be solutions to the equation $\det \rho=0$
also outside the set of density matrices. The boundary of the set of
density matrices can be identified as the {\em closed} surface,
defined by $\det \rho=0$, that encloses the maximally mixed state and
is closest to this point.

%%%%%%%%%%
\subsection{More general transformations}
%%%%%%%%%%

We shall later make use of the complex extension of the
transformations~\pref{unitar}, by allowing $\zeta_a$ to be complex.
This means that the transformation group is extended from $SU(n)$ to
$SL(n,\mathbbm C)$ (the normalization condition $\det U=1$, or
$\zeta_0=0$, is trivial). Transformations of the form
$\tilde{\rho}=V\rho V^{\dag}$ do not respect the trace condition
$\Tr\rho=1$ if $V$ is non-unitary, but they do respect the positivity
condition, because they are still vector transformations of the
form~\pref{conmatU}, with $\ket{\tilde{\psi}_k}=V\ket{\psi_k}$.  This
means that they leave the sector of non-normalized density matrices
invariant. They no longer keep the entropy unchanged, however. Thus,
the larger group $SL(n,\mathbbm C)$ connects a larger set of density
matrices than the restricted group $SU(n)$.

One further generalization is possible.  In fact, even if we allow $V$
to be {\em antilinear}, the transformation $\tilde{\rho}=V\rho V^{\dag}$
still preserves positivity, because Eq.~\pref{conmatU} still holds
with $\ket{\tilde{\psi}_k}=V\ket{\psi_k}$.  This point needs some
elaboration.

An operator $V$ is antilinear if
\be{Vantilin}
V\,(a\,\ket{\psi}+b\,\ket{\phi})
=a^{\ast}V\,\ket{\psi}+b^{\ast}V\,\ket{\phi}
\ee
for any vectors $\ket{\psi}$, $\ket{\phi}$ and complex numbers $a$,
$b$.  Let $\{\ket{i}\}$ be a set of orthonormal basis vectors, let
$\psi_i=\braket{i}{\psi}$ and write $\ket{V\psi}$ for the vector
$V\,\ket{\psi}$.  Then
\be{app02}
\ket{V\psi}
=V\sum_i\psi_i\,\ket{i}
=\sum_{i,j}V_{ji}\psi_i^{\ast}\,\ket{j}
\ee
with $V_{ji}=\bra{j}V\ket{i}$.  
The hermitian conjugate $V^{\dag}$ is
defined in a basis independent way by the identity
\be{antilinhermitconj}
\braket{\psi}{V^{\dag}\phi}\equiv\braket{\phi}{V\psi}
=\sum_{i,j}\phi_j^{\ast}V_{ji}\psi_i^{\ast}
\ee
or equivalently,
\be{app03}
\ket{V^{\dag}\phi}=V^{\dag}\,\ket{\phi}=
\sum_{i,j}\phi_j^{\ast}V_{ji}\,\ket{i}
\ee
By definition, $V$ is {\em antiunitary} if $V^{\dag}=V^{-1}$.

Familiar relations valid for linear operators are not always valid for
antilinear operators.  For example, when $V$ is antilinear and
$\ket{V\psi}=V\ket{\psi}$, it is no longer true that
$\bra{V\psi}=\bra{\psi}V^{\dag}$.  This relation cannot hold, simply
because $\bra{V\psi}$ is a linear functional on the Hilbert space,
whereas $\bra{\psi}V^{\dag}$ is an antilinear functional.  What is
nevertheless true is that
$V\ket{\psi}\bra{\psi}V^{\dag}=\ket{V\psi}\bra{V\psi}\;$.  In fact,
both of these operators are linear, and they act on the vector
$\ket{\phi}=\sum_j\phi_j\ket{j}$ as follows,
\be{app04}
V\ket{\psi}\bra{\psi}V^{\dag}\ket{\phi}
=V\left(\sum_{i,j}\phi_j^{\ast}V_{ji}\psi_i^{\ast}\ket{\psi}\right)
=\sum_{i,j}\psi_iV_{ji}^{\ast}\phi_jV\,\ket{\psi}
=\ket{V\psi}\braket{V\psi}{\phi}
\ee
As a consequence of this identity the form \pref{conmatU} is valid 
for the antiunitary 
transformations, and the positivity is thus 
preserved.

The transposition of matrices, $\rho\ra\rho^T$, obviously preserves
positivity, since it preserves the set of eigenvalues.  This is not an
$SU(n)$ transformation of the form~\pref{adjoint}, as one can easily
check.  However, transposition of a hermitian matrix is the same as
complex conjugation of the matrix, and if we introduce the complex
conjugation operator $K$, which is antilinear and antiunitary, we may
write
\be{transpvectrans}
\rho^T=K\rho K^{\dag}
\ee
Note that transposition is a basis dependent operation.  The complex
conjugation operator $K$ is also basis dependent, it is defined to be
antilinear and to leave the basis vectors invariant,
$K\,\ket{i}=\ket{i}$.  We see that $K^{\dag}=K=K^{-1}$.

One may ask the general question, which are the transformations that
preserve positivity of hermitian matrices.  If we consider an
{\em invertible} linear transformation on the real vector space of
matrices, then it has to be a one-to-one mapping of the extremal
points of the convex set of positive matrices onto the extremal
points.  In other words, it is a one-to-one mapping of one dimensional
projections onto one dimensional projections.  In yet other words, it
is an invertible vector transformation
$\ket{\psi}\ra V\ket{\psi}$, defined up to a phase factor, or more
generally an arbitrary non-zero complex factor, for each pure state
$\ket{\psi}$.  One can show that these complex factors can be chosen
in such a way that $V$ becomes either linear or antilinear, and that
the matrix transformation is $\rho\ra V\rho V^{\dag}$.  However, we
will not go into details about this point here.

%%%%%%%%%%
\subsection{Geometry and separability}
%%%%%%%%%%
We consider next a composite system with two subsystems $A$ and $B$,
of dimensions $n_A$ and $n_B$ respectively.  By definition, the {\em
   separable states} of the system are described by density matrices
that can be written in the form
\be{sep}
\rho=\sum_k\,p_k\, \rho^A_k \otimes\rho^B_k
\ee
where $\rho^A_k$ and $\rho^B_k$ are density matrices of the two
subsystems and $p_k$ is a probability distribution over the set of
product density matrices labelled by $k$. The separable states form a
convex subset of the set of all density matrices of the composite
system, with the {\em pure product states} $\ket{\psi}\bra{\psi}$,
where $\ket{\psi}=\ket{\phi}\otimes\ket{\chi}$, as extremal points. Our
interest is to study the geometry of this set, and thereby the
geometry of the set of entangled states, defined as the complement of
the set of separable states within the full set of density matrices.

The Peres criterion \cite{Peres96} gives a necessary condition
for a density matrix to be separable.  Let us introduce orthonormal
basis vectors $\ket{i}_A$ in ${\cal H}_A$ and $\ket{j}_B$ in ${\cal
H}_B$, as well as the product vectors
\be{eq010a}
\ket{ij}=\ket{i}_A\otimes\ket{j}_B
\ee
We write the matrix elements of the density matrix $\rho$ as
\be{rhomatel}
\rho_{ij;kl}=\bra{ij}\rho\ket{kl}
\ee
The partial transposition with respect to the $B$ system is defined as
the transformation
\be{ptrans}
\rho\ra\rho^P: \quad\rho^P_{ij;kl}\equiv\rho_{il;kj}
\ee
This operation preserves the trace, but not necessarily the positivity
of $\rho$. However, for separable states one can see from the
expansion \pref{sep} that it preserves positivity, because it is just
a transposition of the density matrices $\rho^B_k$ of the subsystem
$B$.

Thus, the Peres criterion states that preservation of positivity under
a partial transposition is a necessary condition for a density matrix
$\rho$ to be separable. Conversely, if the partial transpose $\rho^P$
is {\em not} positive definite, it follows that $\rho$ is
non-separable or entangled. The opposite is not true: if $\rho^P$ is
positive, the density matrix $\rho$ is not necessarily separable.

It should be emphasized that the Peres condition, \ie positivity of
both $\rho$ and $\rho^P$, is independent of the choice of basis
vectors $\ket{i}_A$ and $\ket{j}_B$.  In fact, a change of basis may
result in another definition of the partial transpose $\rho^P$, which
differs from the first one by a unitary transformation, but this does
not change the eigenvalue spectrum of $\rho^P$.  The condition is also
the same if transposition is defined with respect to subsystem $A$
rather than $B$. This is obvious, since partial transposition with
respect to the $A$ subsystem is just the combined transformation
$\rho\ra(\rho^T)^P$.

Let us consider the Peres condition from a geometrical point of view.
We first consider the transposition of matrices, $\rho\ra\rho^T$, and
note that it leaves the Hilbert--Schmidt metric invariant. Being its
own inverse, transposition is an inversion in the space of density
matrices, or a rotation if the number of inverted directions,
$n(n-1)/2$ in an $n\times n$ hermitian matrix, is even.  Since $\rho$
and $\rho^T$ have the same set of eigenvalues, transposition preserves
positivity and maps the set of density matrices onto itself. Thus, the
set of density matrices $\cD$ is invariant under transposition as well
as under unitary transformations.

Similarly, a partial transposition $\rho\ra\rho^P$ preserves the
metric and therefore also corresponds to an inversion or rotation in
the space of matrices. On the other hand, it does not preserve the
eigenvalues and therefore in general does not preserve positivity.
This means that the set of density matrices, $\cD$, is not invariant
under partial transposition, but is mapped into an inverted or rotated
copy $\cD^P$. These two sets will partly overlap, in particular, they
will overlap in a neighbourhood around the maximally mixed state,
since this particular state is invariant under partial transposition.
We note that, even though partial transposition is basis dependent,
the set of transposed matrices $\cD^P$ does not depend on the chosen
basis.  Nor does it depend on whether partial transposition is defined
with respect to subsystem $A$ or $B$.

To sum up the situation, we consider the following three convex sets.
$\cD$ is the full set of density matrices of the composite system,
while $\cP=\cD\cap\cD^P$ is the subset of density matrices that
satisfy the Peres condition, and $\cS$ is the set of separable density
matrices. In general we then have the following inclusions,
\be{sets}
\cS\subset\cP\subset\cD
\ee

The Peres criterion is useful thanks to the remarkable fact that
partial transposition does not preserve positivity.  This fact is
indeed remarkable, for the following reason.  We have seen that any
linear or antilinear vector transformation
$\ket{\psi}\ra V\ket{\psi}$
will preserve the positivity of hermitian matrices by the
transformation
$\rho\ra V\rho V^{\dag}$.  It would seem that
$V=\iden_A\otimes K_B$, a complex conjugation on subsystem $B$, would
be a vector transformation such that $\rho^P=V\rho V^{\dag}$, and
hence, that partial transposition would preserve positivity.  What is
wrong with this argument is that there exists no such operator as
$\iden_A\otimes K_B$.  To see why, choose a complex number $c$ with
$|c|=1$, and consider the transformation of a product vector
$\ket{\psi}=\ket{\phi}\otimes\ket{\chi}$,
\be{ItensorK}
(\iden_A\otimes K_B)\ket{\psi}
=(\iden_A\otimes K_B)((c\,\ket{\phi})\otimes(c^{\ast}\,\ket{\chi}))
=c^2(\ket{\phi}\otimes(K\ket{\chi}))
\ee
The arbitrary phase factor $c^2$ invalidates the attempted definition.

The boundary of the congruent (or reflected) image $\cD^P$ of $\cD$ is
determined by the condition $\det\rho^P=0$ in the same way as the
boundary of the set of density matrices $\cD$ is determined by
$\det\rho=0$. As a consequence, to determine whether a density matrix
$\rho$ belongs to the set $\cP$ is not a hard problem. One simply
checks whether the determinants of $\hat{\rho}$ and $\hat{\rho}^P$ are
both positive for every $\hat{\rho}$ on the line segment between $\rho$
and the maximally mixed state $\rho_0$. However, to check whether a
density matrix is separable and thus belongs to the subset $\cS$ is in
general not easy, even though the definition~\pref{sep} of
separability has a simple form. The exceptional cases are the systems
of Hilbert space dimensions $2\times 2$, $2\times 3$ or $3\times 2$,
where $\cS=\cP$.

%%%%%%%%%%
\subsection{Schmidt decomposition and transformation to a 
standard form}
%%%%%%%%%%

A general density matrix of the composite system can be expanded as
\be{expand00a}
\rho=\sum_{a=0}^{n_A^2-1}\sum_{b=0}^{n_B^2-1}
\xi_{ab}\,J^A_a\otimes  J^B_b
\ee
where the coefficients $\xi_{ab}$ are real, and $J^A_a$ and $J^B_b$
are orthonormal basis vectors of the two subsystems.  We may use our
convention that $J^A_0=\iden_A/\sqrt{n_A}$ and
$J^B_0=\iden_B/\sqrt{n_B}$, and that $J^A_k$ and $J^B_k$ with $k>0$
are generators of $SU(n_A)$ and $SU(n_B)$.

A Schmidt decomposition is a diagonalization of the above expansion.
By a suitable choice of basis vectors $\hat{J}^A_a$ and $\hat{J}^B_b$,
depending on $\rho$, we may always write
\be{expand00b}
\rho=\sum_{a=0}^{n_A^2-1}
\hat{\xi}_a\,\hat{J}^A_a\otimes\hat{J}^B_a
\ee
assuming that $n_A\leq n_B$.  There exist many different such diagonal
representations of a given $\rho$, in fact it is possible to impose
various extra conditions on the new basis vectors.  It is usual to
impose an orthonormality condition, that the new basis vectors should
be orthonormal with respect to some positive definite scalar product.
Then the Schmidt decomposition of $\rho$ is the same as the singular
value decomposition \cite{Golub} of the $n_A^2\times n_B^2$ matrix
$\xi_{ab}$.  Below, we will introduce a Schmidt decomposistion based
on other types of extra conditions.

The usefulness of the representation~\pref{expand00b} is limited by
the fact that we expand in basis vectors depending on $\rho$.
However, we may make a transformation of the form
\be{prodformtransf}
\rho\ra\tilde{\rho}=V\rho V^{\dag}
\ee
where $V=V_A\otimes V_B$ is composed of transformations
$V_A\in SL(n_A,\mathbbm C)$ and $V_B\in SL(n_B,\mathbbm C)$ that act
independently on the two subsystems and transform the basis vectors
$\hat{J}^A_a$ and $\hat{J}^B_b$ into $V_A\hat{J}^A_aV_A^{\dag}$ and
$V_B\hat{J}^B_bV_B^{\dag}$.  A transformation of this form obviously
preserves the set $\cS$ of separable states, since a sum of the
form~\pref{sep} is transformed into a sum of the same form.  It also
preserves the set $\cP$ of density matrices satisfying the Peres
condition.  In fact, it preserves the positivity not only of $\rho$,
but also of the partial transpose $\rho^P$, since
\be{rhoPtransf}
(\tilde{\rho})^P
=(V_A\otimes V^{\ast}_B)\rho^P(V_A\otimes V^{\ast}_B)^{\dag}
\ee
Here $V^{\ast}_B$ is the complex conjugate of $V_B$. What is not
preserved by the transformation is the trace, but this can easily be
corrected by introducing a normalization factor. Such transformations have been 
considered e.g. by Cen et al. \cite{Cen03}.

As we will later show, it is possible to choose the transformation
$V=V_A\otimes V_B$ in such a way that the transformed and normalized
density matrix $\tilde\rho$ can be brought into the special form
\be{schmidt1x}
\tilde{\rho}
={1\over n_An_B}\left(\iden+\sum_{k=1}^{n_A^2-1} \tilde{\xi}_{k}\,
\tilde{J}^A_k\otimes\tilde{J}^B_k\right)
\ee
with $\tilde{J}^A_k$ and $\tilde{J}^B_k$ as new sets of traceless
orthonormal basis vectors.  We have some freedom in choosing $V_A$ and
$V_B$, because the form~\pref{schmidt1x} is preserved by unitary
transformations of the form $U=U_A\otimes U_B$.

In the case $n_A=n_B=2$ we may choose $V_A$ and $V_B$ so that
$\tilde{J}^A_k$ and $\tilde{J}^B_k$ are fixed sets of basis vectors,
independent of the density matrix $\rho$.  In particular, we may use
the standard Pauli matrices as basis vectors.  In this way we define a
special form of the density matrices $\tilde\rho$, which we refer to
as the standard form. Any density matrix can be brought into this form
by a transformation that preserves separability and the Peres
condition.  All matrices of the resulting standard form commute and
can be simultaneously diagonalized.  This makes it easy to prove the
equality $\cS=\cP$, and thereby solve the separability problem.
Although this result is well known, the proof given here is simpler
than the original proof.

The decomposition \pref{schmidt1x} is generally valid, but when either
$n_A$, $n_B$, or both are larger than 2, it is impossible to choose
both $\tilde{J}^A_k$ and $\tilde{J}^B_k$ to be independent of $\rho$.
Simply by counting the number of parameters one easily demonstrates
that this cannot be done in higher dimensions. Thus, the product
transformations $V_A\otimes V_B$ are specified by $2n_A^2+2n_B^2-4$
parameters, while the number of parameters in \pref{schmidt1x} is
$n_A^2-1$, when the generators $\tilde{J}^A_k$ and $\tilde{J}^B_k$ are
fixed.  This gives a total number of parameters $3n_A^2+2n_B^2-5$,
when $n_A\leq n_B$, compared to the number of parameters of the
general density matrix, which is $n_A^2 n_B^2-1$.  Only for
$n_A=n_B=2$ do these numbers match.

The mismatch in the number of parameters shows that the independent
transformations $V_A$ and $V_B$ performed on the two subsystems are
less efficient in simplifying the form of the density matrices in
higher dimensions.  In particular, it is impossible to transform all
density matrices to a standard form of commuting matrices.  Thus, the
question of separability is no longer trivially solved.  Nevertheless,
we consider the Schmidt decomposition to be interesting and important,
if only because the number of dimensions in the problem is reduced.
We expect this to be useful, even if, at the end, separability can
only be determined by numerical methods.

%%%%%%%%%%
\section{The two-level system}
%%%%%%%%%%

The density matrices of a two-level system describe the states of a
qubit, and represent a simple, but important, special case. It is well
known that the normalized density matrices, expressed in terms of the
Pauli matrices $\bs\sigma=(\sigma_1,\sigma_2,\sigma_3)$ as
\be{twodensity}
\rho=\half\,(\iden+\v r\cdot\bs\sigma)
\ee
can geometrically be pictured as the interior of a three dimensional
unit sphere, the {\em Bloch sphere}, with each point identified by a
vector $\v r$. The two-level case is special in that the pure states
form the complete surface $|\v r|=1$ of the set of density matrices.
The diagonal $2\times 2$ matrices, in any chosen basis, correspond to
a line segment through the origin with the two pure basis states as
end points.

The two-level system is special also in the sense that the Euclidean
metric of the three dimensional space of density matrices can be
extended in a natural way to an indefinite metric in four dimensions.
The extension is analogous to the extension from the three dimensional
Euclidean metric to the four dimensional Lorentz metric in special
relativity. Since it is useful for the discussion of entanglement in
the two qubit system, we shall briefly discussed it here.

\begin{figure}[h]
\begin{center}
\includegraphics[width=7cm]{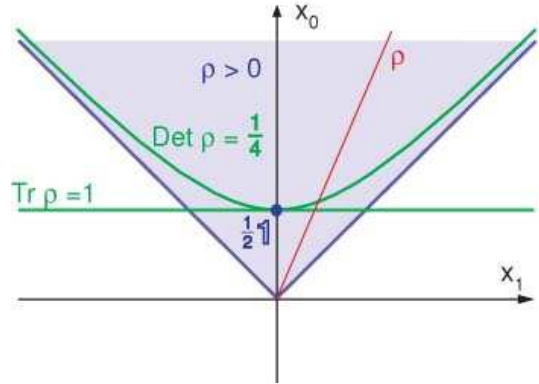}
\end{center}
\caption{{\small A relativistic view of the density matrices of a
    two-level system. The positive hermitian matrices are represented
    by the forward light cone and a density matrix by a time-like or
    light-like ray (red line). The standard normalization $\Tr
    \rho=x^0=1$ breaks relativistic invariance (green horizontal
    line), but may be replaced by the relativistically invariant
    normalization $\det\rho=x^{\mu}x_{\mu}/4=1/4$ (green hyperbola). }
}
\label{lightcone}
\end{figure}

We write the density matrix in relativistic notation as
\be{twodensity2}
\rho=\half\,x^{\mu}\sigma_{\mu}
\ee
where $\sigma_0$ is the identity matrix $\iden$. The trace
normalization condition is that $\Tr\rho=x^0=1$. We may relax this
condition and retain only the positivity condition on $\rho$, which
means that $x^0$ is positive and dominates the vector part $\v r$ of
the four-vector $x^{\mu}$, as expressed by the covariant condition
\be{pos}
4\,\det\rho=(x^0)^2-|\v r|^2=x^{\mu}x_{\mu}=g_{\mu\nu}x^{\mu}x^{\nu}
\geq 0
\ee
In other words, the four-vector $x^{\mu}$ is restricted by the
positivity condition to be either a time-like vector inside the
forward light cone, or a light-like vector on the forward light cone.
The light-like vectors correspond to pure states, and the time-like
vectors to mixed states.  As already discussed, all points on a given
line through the origin represent the same normalized density matrix
(see Fig.~\ref{lightcone}).

%%%%%%%%%%%%
%%%%%%%%%%%%

Positivity is conserved by matrix transformations of the form
\be{vtrans}
\rho\ra\tilde{\rho}=V\rho V^{\dag}
=\half\,L^{\nu}_{\;\;\mu}x^{\mu}\sigma_{\nu}
\ee
If we restrict $V$ to be linear (not antilinear), invertible, and
normalized by $\det V=1$, then it belongs to the group
$SL(2,\mathbbm C)$, and $L$ is a continuous Lorentz transformation
(continuous in the sense that it can be obtained as a product of
small transformations).  Thus, preservation of positivity by the
$SL(2,\mathbbm C)$ transformations corresponds to preservation of the
forward light cone by the continuous Lorentz transformations.

In order to compare the indefinite Lorentz metric to the Euclidean
scalar product $\Tr(AB)$ introduced earlier, we introduce the
operation of space inversion,
\be{adjoint1}
\sigma_{\mu}=(\iden,\bs\sigma)\ra
\bar \sigma_{\mu}\equiv (\iden,-\bs \sigma)
\ee
It is obtained as a combination of matrix transposition, or
equivalently complex conjugation, which for the standard Pauli
matrices inverts the sign of $\sigma_y=\sigma_2$, and a rotation of
$\pi$ about the $y$-axis.  Thus, for a general hermitian $2\times 2$
matrix $A$ it acts as
\be{Atrans}
A\ra\bar A=RA^TR^{\dag}
\ee
where
\be{Rypi}
R=\left(\begin{array}{cc}
0 & -1 \\
1 & 0
\end{array}\right)
\ee
We may also write $A^T=KAK^{\dag}$, where $K=K^{\dag}$ is
complex conjugation.

The Lorentz metric is now expressed by
\be{metric}
\Tr(\bar\sigma_{\mu}\sigma_{\nu})=2g_{\mu\nu}
\ee
and the Lorentz invariant scalar product between two hermitian
matrices $A=a^{\mu}\sigma_{\mu}$ and $B=b^{\mu}\sigma_{\mu}$ is
\be{scalprod1}
\Tr(\bar A B)=2a^{\mu}b_{\mu}
\ee
The invariance of the scalar product \pref{metric} can be seen
directly from the transformation properties under $SL(2,\mathbbm C)$
transformations,
\be{sltrans}
A\ra A'=VAV^{\dag} \Rightarrow \bar A\ra \bar A'=V^{\dag -1}\bar A V^{-1}
\ee
Thus, $A$ and $\bar A$ transform under contragredient representations
of $SL(2,\mathbbm C)$.

The Lorentz transformed Pauli matrices
\be{Lorentz}
\tilde\sigma_{\mu}
=V\sigma_{\mu}V^{\dag}=L^{\nu}_{\;\mu} \sigma_{\nu}\;,\quad
V \in SL(2,\mathbbm C)
\ee
satisfy the same metric condition \pref{metric} as the standard Pauli
matrices $\sigma_{\mu}$. Conversely, any set of matrices
$\tilde\sigma_{\mu}$ satisfying relations of the form \pref{metric}
are related to the Pauli matrices by a Lorentz transformation, which
need not, however, be restricted to the continuous part of the Lorentz
group, but may include space inversion or time reversal, or both.

It is clear from the relativistic representation that any density
matrix can be reached from the maximally mixed state $\rho_0=\iden/2$
by a transformation corresponding to a boost. The Lorentz boosts
generate from $\rho_0$ a three dimensional hyperbolic surface (a
``mass shell'') where $x^{\mu}x_{\mu}=1$.  This surface will intersect
any time-like line once and only once. Thus, any mixed state is
obtained from $\rho_0$ by a unique boost.  However, the pure states,
corresponding to light-like vectors, can only be reached
asymptotically, when the boost velocity approaches the speed of light,
here set equal to 1. The form of the $SL(2,\mathbbm C)$ transformation
corresponding to a pure boost is
\be{boost}
V(\bs \xi)=\exp\!\left(\half\,\bs \xi\cdot \bs \sigma\right)
\ee
where the three dimensional real vector $\bs \xi$ is the boost
parameter, called rapidity.  Since the boost matrices are hermitian, a
density matrix defined by a boost of the maximally mixed state will
have the form
\be{bdens1}
\rho=N(\xi)V(\bs \xi)^2=N( \xi)(\cosh\xi\,\iden
+\sinh\xi\;\hat{\bs \xi}\cdot\bs\sigma)\;,\quad
\hat{\bs \xi}=\frac{\bs\xi}{\xi}
\ee
where $N( \xi)$ is a normalization factor determined by the trace
normalization condition. The normalized density matrix is
\be{bdens2}
\rho=\half\,(\iden+\tanh\xi\;\hat{\bs \xi}\cdot\bs\sigma)
\ee

Thus, the boost parameter $\bs \xi$ gives a representation which is an
alternative to the Bloch sphere representation. The relation between
the parameters $\v r$ and $\bs\xi$ is that
\be{rel}
\v r=\tanh\xi\;\hat{\bs\xi}
\ee
which means that $\v r$ can be identified as the velocity of the
boost, in the relativistic picture, with $|\v r|=1$ corresponding to
the speed of light.  We note that the positivity condition gives no
restriction on $\bs \xi$, and the extremal points, \ie\ the pure
states, are points at infinity in the $\bs \xi$ variable.

%%%%%%%%%%
\section{Entanglement in the $2\times 2$ system}
%%%%%%%%%%

We consider now in some detail entanglement between two two-level
systems.  We will show that with the use of non-unitary
transformations the density matrices can be written in a standardized
Schmidt decomposed form.  In this form the question of separability is
easily determined and the equality of the two sets $\cP$ and $\cS$ is
readily demonstrated.

%%%%%%%%%%
\subsection{Schmidt decomposition by Lorentz transformations}
%%%%%%%%%%

We consider transformations $V=V_A\otimes V_B$ composed of
$SL(2,\mathbbm C)$ transformations acting independently on the two
subsystems, and therefore respecting the product form \pref{sep} of
the separable matrices.  We will show that by means of such
transformations any density matrix of the composite system can be
transformed to the form
\be{stanfor}
\tilde{\rho}
=\frac{1}{4}\left(\iden+\sum_{k=1}^3 d_k\,\sigma_k\otimes \sigma_k\right)
\ee
which we refer to as the standard form.  Note that the real
coefficients $d_k$ must be allowed to take both positive and negative
values.

We start by writing a general density matrix in the form
\be{denmat}
\rho=c^{\mu\nu}\,\sigma_{\mu}\otimes\sigma_{\nu}
\ee
The transformation $V=V_A\otimes V_B$ produces independent Lorentz
transformation $L_A$ and $L_B$ on the two subsystems,
\be{denmat01}
\rho\ra\tilde{\rho}=V\rho V^{\dag}
=\tilde{c}^{\,\mu\nu}\sigma_{\mu}\otimes\sigma_{\nu}
\ee
with
\be{denmat02}
\tilde{c}^{\,\mu\nu}=L^{\mu}_{A\rho}L^{\nu}_{B\sigma}c^{\rho\sigma}
\ee
The Schmidt decomposition consists in choosing $L_A$ and $L_B$ in such
a way that $\tilde{c}^{\,\mu\nu}$ becomes diagonal.  We will show that
this is always possible when $\rho$ is strictly positive.

Note that the standard Schmidt decomposition, also called the singular
value decomposition, involves a compact group of rotations, leaving
invariant a positive definite scalar product.  The present case is
different, because it involves a non-compact group of
Lorentz-transformations, leaving invariant an indefinite scalar
product.

The positivity condition on $\rho$ plays an essential part in the
proof.  It states that
\be{posdens}
\bra{\psi}\rho\ket{\psi}\geq 0
\ee
where $\ket{\psi}$ is an arbitrary state vector. Let us consider a
density matrix $\rho$ which is {\em strictly} positive so that
\pref{posdens} is satisfied with $>$ and not only with $\geq$, and let
us restrict $\ket{\psi}$ to be of product form,
$\ket{\psi}=\ket{\phi}\otimes\ket{\chi}$. With $\rho$ expressed by
\pref{denmat}, the positivity condition then implies
\be{poscmat}
c^{\mu\nu}m_{\mu}n_{\nu}>0
\ee
with
\be{nAB}
m_{\mu}=\bra{\phi}\sigma_{\mu}\ket{\phi}\;,\quad
n_{\nu}=\bra{\chi}\sigma_{\nu}\ket{\chi}
\ee
These two four-vectors are on the forward light cone, in fact, it is
easy to show that
\be{lightlike}
m_{\mu}m^{\mu}=n_{\mu}n^{\mu}=0\;,\quad m_0=n_0=1
\ee
We note that by varying the state vectors $\ket{\phi}$ and
$\ket{\chi}$ all directions on the light cone can be reached.  The
inequality~\pref{poscmat} holds for forward time-like vectors as well,
because any such vector may be written as a linear combination of two
forward light-like vectors, with positive coefficients.  We may
actually write a stronger inequality
\be{poscmat03}
c^{\mu\nu}m_{\mu}n_{\nu}\geq C>0
\ee
valid for all time-like or light-like vectors $m$ and $n$ with
$m_0=n_0=1$. In fact, this inequality holds because the set of such
pairs of four-vectors $(m,n)$ is compact.

Now define the function
\be{funcmn}
f(m,n)={c^{\mu\nu}m_{\mu}n_{\nu}\over
\sqrt{m^{\rho}m_{\rho}\,n^{\sigma}n_{\sigma}}}
\ee
It is constant for $m$ and $n$ lying on two fixed, one dimensional rays
inside the forward light cone. It goes to infinity when either $m$ or
$n$ becomes light-like, because
\be{funcmn01}
f(m,n)\geq{Cm_0n_0\over
\sqrt{m^{\rho}m_{\rho}\,n^{\sigma}n_{\sigma}}}
\ee
Using again a compactness argument, we conclude that there exist
four-vectors $\bar{m}$ and $\bar{n}$ such that $f(\bar{m},\bar{n})$ is
minimal. We may now choose the Lorentz transformations $L_A$ and $L_B$
such that
\be{LALBdef}
L^0_{A\mu}=\bar{m}_{\mu}\;,\quad
L^0_{B\mu}=\bar{n}_{\mu}
\ee
assuming the normalization conditions
$\bar{m}^{\mu}\bar{m}_{\mu}=\bar{n}^{\mu}\bar{n}_{\mu}=1$.  This
defines $L_A$ and $L_B$ uniquely up to arbitrary three dimensional
rotations. Define
\be{funcmn02}
\tilde{f}(m,n)={\tilde{c}^{\,\mu\nu}m_{\mu}n_{\nu}\over
\sqrt{m^{\rho}m_{\rho}\,n^{\sigma}n_{\sigma}}}
\ee
Since $\tilde{f}(m,n)=f(\tilde{m},\tilde{n})$, with
$\tilde{m}_{\mu}=L^{\rho}_{A\mu}m_{\rho}$ and
$\tilde{n}_{\mu}=L^{\rho}_{B\mu}n_{\rho}$, it follows that
$\tilde{f}(m,n)$ has a minimum at $m=n=(1,0,0,0)$.
The condition for an extremum at $m=n=(1,0,0,0)$ is that
$\tilde{c}^{\,0k}=\tilde{c}^{\,k0}=0$ for $k=1,2,3$, so that
\be{denmat04}
\tilde{\rho}
=\tilde{c}^{\,00}\,\iden+\sum_{k=1}^3\sum_{l=1}^3
\tilde{c}^{\,kl}\,\sigma_k\otimes\sigma_l
\ee
The coefficient $\tilde{c}^{\,00}$ is the minimum value of $f(m,n)$,
and hence positive.

The last term of Eq. \pref{denmat04} can be
diagonalized by a standard Schmidt decomposition,
and by a further normalization $\tilde\rho$ can
be brought into the form \pref{schmidt1x}.
Finally, a unitary transformation
$\tilde{\rho}\ra U\tilde{\rho}U^{\dag}$ of the product form
$U=U_A\otimes U_B$ may be performed, where the
unitary matrices $U_A$ and $U_B$ may be chosen so
that
$U_A\tilde{J}^A_kU_A^{\dag}=\pm\sigma_k$ and
$U_B\tilde{J}^B_kU_B^{\dag}=\pm\sigma_k$.  This is aways possible,
because $SU(2)$ transformations generate the full three dimensional
rotation group, excluding inversions.  In this way we obtain the
standard form~\pref{stanfor}.

Note that the standard form ~\pref{stanfor} of a given density matrix
$\rho$ is not unique, because there exists a discrete subgroup of 24
unitary transformations that transform one matrix of this form into
other matrices of the same form.  This group includes all permutations
of the three basis vectors $\sigma_k\otimes\sigma_k$, as well as
simultaneous reversals of any two of the basis vectors.  It is the
full symmetry group of a regular tetrahedron.  If we want to make the
standard form unique we may, for example, impose the conditions
$d_1\geq d_2\geq|d_3|$, allowing both positive and negative values of
$d_3$.

%%%%%%%%%%
\subsection{Density matrices in the standard form}
%%%%%%%%%%

The density matrices of the standard form \pref{stanfor} define a
convex subset of lower dimension than the full set of density
matrices. It is a three dimensional section of the 15~dimensional set
of density matrices, consisting of commuting (simultaneously
diagonalizable) matrices. The eigenvalues, as functions of the
parameters $d_k$ of Eq.~\pref{stanfor}, are
\be{eigenvalsd1d2d3}
\lambda_{1,2}={1\over 4}\,(1\pm(d_1-d_2)+d_3)\;, \quad
\lambda_{3,4}={1\over 4}\,(1\pm(d_1+d_2)-d_3)
\ee
The pure states (with eigenvalues $1,0,0,0$) that are the extremal
points of the convex set of commuting matrices are specified by the
conditions
\be{Bell}
|d_1|=|d_2|=|d_3|=1\;,\quad d_1d_2d_3=-1
\ee
There are four such states, corresponding to the four corners of the
tetrahedron of diagonal density matrices, and these are readily
identified as four orthogonal Bell states (maximally entangled pure
states).

\begin{figure}[h]
\begin{center}
\includegraphics[width=6cm]{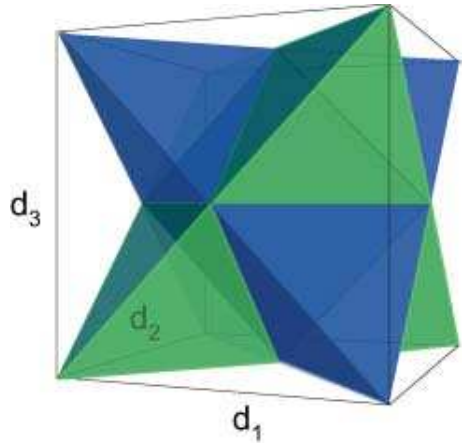}
\end{center}
\caption{{\small Geometric representation of the diagonal matrices
     spanned by orthogonal Bell states of the $2\times 2$ system. The
     density matrices form a tetrahedron (green) while the matrices
     obtained from this by a partial transposition defines a mirror
     image (blue). The separable states are defined by the intersection
     between the two tetrahedra, and forms an octahedron (in the center).} }
\label{Polygon}
\end{figure}

We now consider the action of the partial transposition on the
tetrahedron of diagonal density matrices. It preserves the standard
form, and transforms the coefficients as $d_1\ra d_1, d_2\ra -d_2,
d_3\ra d_3$. Thus it produces a mirror image of the tetrahedron, by a
reflection in the $d_2$ direction (see Fig.~\ref{Polygon}). The density
matrices of standard form belonging to the set $\cP$, \ie satisfying
the Peres condition that they remain positive after the partial
transposition, form an octahedron which is the intersection of the two
tetrahedra.

%%%%%%%%%%%%
%%%%%%%%%%%%

We will now show that for the density matrices of standard form the
Peres condition is both necessary and sufficient for separability.
What we have to show is that all the density matrices of the
octahedron are separable. Since, in general, the separable matrices
form a convex set, it is sufficient to show that the corners of the
octahedron correspond to separable states.

The density matrices of the octahedron satisfy a single inequality
\be{octa}
|d_1|+|d_2|+|d_3|\leq 1
\ee
and its six corners are $(d_1,d_2,d_3)=(\pm1,0,0),
(0,\pm1,0),(0,0,\pm1)$, corresponding to the midpoints of the six
edges of each of the two tetrahedra.  The corners are separable by the
identities
\be{corner}
\frac{1}{4}\,(\iden\pm\sigma_k\otimes\sigma_k)
={1\over 8}\,
[(\iden+\sigma_k)\otimes(\iden\pm\sigma_k)
+(\iden-\sigma_k)\otimes(\iden\mp\sigma_k)]
\ee

This completes our proof that the Peres condition is both necessary
and sufficient for separability of density matrices on the standard
form.  Furthermore, since any (non-singular) density matrix can be
obtained from a density matrix of standard form by a transformation
that preserves both separability and the Peres condition, this
reproduces the known result that for the $2\times 2$ system the set of
density matrices that remain positive after a partial transposition is
identical to the set of separable density matrices.

With this we conclude the discussion of the two-qubit system. The main
point has been to show the usefulness of applying the non-unitary
Lorentz transformations in the discussion of separability. Also in
higher dimensions such transformations can be applied in the form of
$SL(n,\mathbbm C)$ transformations, although not in precisely the same
form as with two-level systems.

%%%%%%%%%
\section{Higher dimensions}
%%%%%%%%%%

The relativistic formulation is specific for the two-level system, but
some elements can be generalized to higher dimensions. We consider
first a single system with Hilbert space dimension $n$. Again, if the
trace condition is relaxed, the symmetry group $SU(n)$ of the set of
density matrices is extended to $SL(n,\mathbbm C)$.  The
Hilbert--Schmidt metric is {\em not} invariant under this larger
group, but the determinant is invariant for any $n$.  However, it is
only for $n=2$ that the determinant is quadratic and can be
interpreted as defining an invariant indefinite metric.

The generalization to $n>2$ of the Lorentz boosts are the hermitian
matrices
\be{Vtrans}
V=V^{\dag}\;,\quad \det V=1
\ee
and expressed in terms of the $SU(n)$ generators $\v J$ they have the
form
\be{exp}
V=\exp(\bs\xi\cdot\v J)
\ee
with real group parameters $\bs\xi$.  Here $\v J$ and $\bs\xi$ are
$n^2-1$ dimensional vectors.  Any {\em strictly positive} density
matrix $\rho$ (with $\det\rho>0$) can be factorized in terms of
hermitian matrices \pref{exp} as
\be{densboost}
\rho=NV^2=N\,\exp(2\bs\xi \cdot \v J)
\ee
with the normalization factor
\be{norm}
N={1\over\Tr\exp(2\bs\xi \cdot \v J)}
=(\det\rho)^{{1\over n}}
\ee
Thus, in the same way as for $n=2$, any strictly positive density
matrix can be generated from the maximally mixed state by an
$SL(n,\mathbbm C)$ transformation of the form \pref{exp}. The
boundary matrices, however, which satisfy $\det\rho=0$, cannot be
expressed in this way, they can be reached by hermitian
transformations only asymptotically, as $|\bs\xi|\ra\infty$.  In this
limit $N\ra0$, and therefore $\Tr V^2\ra\infty$ for the non-normalized
density matrix $V^2$.

%%%%%%%%%%
\subsection{Schmidt decomposition}
%%%%%%%%%%

We now consider a composite system, consisting of two subsystems of
dimension $n_A$ and $n_B$, and assume $\rho$ to be a strictly positive
density matrix on the Hilbert space
${\cal H}={\cal H}_A\otimes{\cal H}_B$ of the composite system.  The
general expansion of $\rho$ in terms of the $SU(n)$ generators is
given by \pref{expand00a}. Our objective is to show that by a
transformation of the form~\pref{prodformtransf}, with
$V_A\in SL(n_A,\mathbbm C)$ and $V_B\in SL(n_B,\mathbbm C)$,
followed by normalization, the density matrix can be transformed to
the simpler form~\pref{schmidt1x}.

Let
${\cal D}_A$ and ${\cal D}_B$ be the sets of density matrices of the
two subsystems $A$ and $B$.  The Cartesian product
${\cal D}_A\times{\cal D}_B$, consisting of all product density
matrices $\rho_A\otimes\rho_B$ with normalization
$\Tr\rho_A=\Tr\rho_B=1$, is a compact set of matrices on the full
Hilbert space ${\cal H}$.  For the given density matrix $\rho$ we
define the following function of $\rho_A$ and $\rho_B$, which does not
depend on the normalizations of $\rho_A$ and $\rho_B$,
\be{ffunc2}
f(\rho_A, \rho_B)={\Tr\left(\rho\,(\rho_A\otimes\rho_B)\right)\over
(\det\rho_A)^{1/n_A}\,(\det\rho_B)^{1/n_B}}
\ee
This function is well defined on the interior of
${\cal D}_A\times{\cal D}_B$, where $\det\rho_A>0$ and
$\det\rho_B>0$.  Because $\rho$ is assumed to be strictly
positive, we have the strict inequality
\be{eq002}
\Tr\left(\rho\,(\rho_A\otimes\rho_B)\right)>0
\ee
and since ${\cal D}_A\times{\cal D}_B$ is compact, we have an even
stronger inequality on ${\cal D}_A\times{\cal D}_B$,
\be{eq003}
\Tr(\rho\,(\rho_A\otimes\rho_B))\geq C>0
\ee
with a lower bound $C$ depending on $\rho$.  It follows that
$f\to\infty$ on the boundary of ${\cal D}_A\times{\cal D}_B$, where
either $\det\rho_A=0$ or $\det\rho_B=0$.  It follows further that $f$
has a positive minimum on the interior of
${\cal D}_A\times{\cal D}_B$, with the minimum value attained for at
least one product density matrix $\tau_A\otimes\tau_B$ with
$\det\tau_A>0$ and $\det\tau_B>0$.  For $\tau_A$ and $\tau_B$ we may
use the representation \pref{densboost}, written as
\be{eq007}
\tau_A=T_A^{\dag}T_A\;,\qquad
\tau_B=T_B^{\dag}T_B
\ee
ignoring normalization factors.  The matrices
$T_A\in SL(n_A,\mathbbm C)$ and $T_B\in SL(n_B,\mathbbm C)$ may be
chosen to be hermitian, but they need not be, since they may be
multiplied from the left by arbitrary unitary matrices.  We further
write $T=T_A\otimes T_B$, so that
\be{eq008}
\tau_A\otimes\tau_B=T^{\dag}T
\ee
Now define a transformed density matrix
\be{transrho}
\tilde{\rho}=T\rho T^{\dag}
\ee
and define
\be{ffunc3}
\tilde{f}(\rho_A,\rho_B)
={\Tr\left(\tilde{\rho}\,(\rho_A\otimes\rho_B)\right)\over
(\det\rho_A)^{1/n_A}\,(\det\rho_B)^{1/n_B}}
=f(T_A^{\dag}\rho_AT_A\,,\,T_B^{\dag}\rho_BT_B)
\ee
This transformed function $\tilde{f}$ has a minimum for
\be{min}
\rho_A\otimes\rho_B=(T^{\dag})^{-1}\tau_A\otimes\tau_BT^{-1}
=\iden_A\otimes\iden_B=\iden
\ee
Since $\tilde{f}$ is stationary under infinitesimal variations about
the minimum, it follows that
\be{minvar}
\Tr(\tilde\rho\;\delta (\rho_A\otimes\rho_B))=0
\ee
for all infinitesimal variations
\be{varrho}
\delta (\rho_A\otimes\rho_B)
=\delta\rho_A\otimes \iden_B+\iden_A\otimes \delta\rho_B
\ee
subject to the constraints
$\det(\iden_A+\delta\rho_A)=\det(\iden_B+\delta\rho_B)=1$, or
equivalently,
\be{constdettr0}
\Tr(\delta\rho_A)=\Tr(\delta\rho_B)=0
\ee
The variations satisfying the constraints are the general linear
combinations of the $SU$ generators,
\be{vargen}
\delta\rho_A=\sum_i\delta c^A_i\,J^A_i\;,\qquad
\delta\rho_B=\sum_j\delta c^B_j\,J^B_j
\ee
It follows that
\be{gentrace}
\Tr(\tilde\rho\;J^A_i\otimes\iden_B)=\Tr(\tilde\rho\;\iden_A\otimes J^B_j)=0
\ee
for all $SU(n_A)$ generators $J^A_i$ and all $SU(n_B)$ generators
$J^B_j$.  This means that the terms proportional to
$J^A_i\otimes\iden_B$ and $\iden_A\otimes J^B_j$ vanish in the
expansion for $\tilde\rho$, which therefore has the form
\be{rhotilexp}
\tilde\rho
={1\over{n_An_B}}\left(\iden+\sum_{k=1}^{n_A^2-1}\sum_{l=1}^{n_B^2-1} 
\xi_{kl} J^A_k\otimes  J^B_l\right)
\ee
In order to write $\tilde{\rho}$ in the Schmidt decomposed form
\pref{schmidt1x}, we have to make a change of basis, from the fixed
basis sets $J^A_i$ and $J^B_j$ to other orthonormal $SU$
generators $\tilde J^A_i$ and $\tilde J^B_j$ depending on
$\tilde{\rho}$.  This final Schmidt decomposition involves a standard
singular value decomposition of the matrix $\xi_{kl}$ by orthogonal
transformations.  We may make further unitary transformations
$U=U_A\otimes U_B$, but as already pointed out, this is in general not
sufficient to obtain a standard form independent of $\rho$.

%%%%%%%%%%
\section{Numerical approach to the study of separability}
%%%%%%%%%%

In higher dimensions the Peres condition is not sufficient to identify
the separable states. In other words, there exist entangled states
that remain positive after a partial transposition. This is known not
only from general theoretical considerations \cite{Horodecki96}, but
also from explicit examples \cite{Horodecki97}. States of this type
have been referred to as having {\em bound} entanglement.  However,
whereas it is a fairly simple task to check the Peres condition, it is
in general difficult to identify the separable states
\cite{Gurvits03}.

In this section we discuss a general numerical method for
identifying separability, previously introduced in \cite{Dahl06}. It 
is based 
on an iterative scheme for calculating
the closest separable state and the distance to it, given an arbitrary
density matrix (test state).  The method can be used to test single
density matrices for separability or to make a systematic search to
identify the boundary of the set of separable density matrices. After
giving an outline of the method we show how to apply the method in a
numerical study of bound entanglement in a $3\times 3$ system.

\subsection{Outline of the method}

\begin{figure}[h]
\begin{center}
\includegraphics[width=5cm]{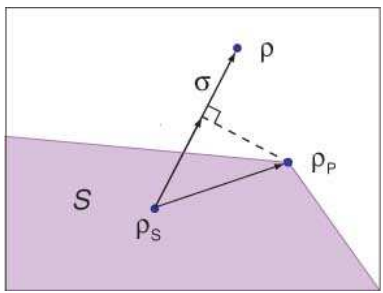}
\end{center}
\caption{{\small Schematic illustration of the method of finding the
    separable state closest to a test matrix $\rho$.  The matrix
    $\rho_s$ represents the best approximation to the closest
    separable state at a given step in the iteration, while $\rho_p$
    is the product matrix that maximizes the scalar product with the
    matrix $\sigma=\rho-\rho_s$. This matrix is used to improve
    $\rho_s$ at the next step in the iteration.  The colored area
    $\cS$ in the figure illustrates a section through the set of
    separable density matrices.  }}
\label{Best}
\end{figure}

Assume a test state $\rho$ has been chosen.  This may typically be
close to the boundary of the set of states that satisfy the Peres
condition.  Let $\rho_s$ be a separable state, an approximation in the
iterative scheme to the closest separable state.  We may start for
example with $\rho_s=\iden/n$, the maximally mixed state, or with any
pure product state.  The direction from $\rho_s$ to $\rho$ is denoted
$\sigma=\rho-\rho_s$.  In order to improve the estimate $\rho_s$ we
look for a pure product state $\rho_p$ that maximizes the scalar
product
\be{eq009}
s=\Tr((\rho_p-\rho_s)\sigma)
\ee
or equivalently, maximizes $s'=\Tr(\rho_p\sigma)$ (see
Fig.~\ref{Best}).
%%%%%%%%%%%%
%%%%%%%%%%%%
If $s>0$, then it is possible to find a closer separable 
state $\rho'_s$
by mixing in the product state $\rho_p$.  This search for closer
separable states is iterated, either until no pure product state
$\rho_p$ can be found such that $s>0$, which means that $\rho_s$ is
already the unique separable state closest to $\rho$, or until some
other convergence criterion is satisfied.

There are two separate mathematical subproblems that have to be solved
numerically in this scheme. The first problem is to find the pure product state
maximizing the scalar product $s'$.  The second problem is the so
called quadratic programming problem: given a finite number of pure
product states, to find the convex combination
of these  which is closest to the given state
$\rho$. Our approach to these two problems is described briefly 
below. We refer 
to reference \cite{Dahl06} for more details.

To approach the first subproblem, note that a pure product state
$\rho_p$ has matrix elements of the form
\be{eq011}
\langle ij|\rho_p | kl\rangle
=\phi_i\chi_j\phi_k^{\ast}\chi_l^{\ast}
\ee
where $\sum_i|\phi_i|^2=\sum_j|\chi_j|^2=1$.  We want to find complex
coefficients $\phi_i$ and $\chi_j$ that maximize
\be{eq012}
s'=\Tr(\rho_p\sigma)=\sum_{i,j,k,l}
\phi_i^{\ast}\,\chi_j^{\ast}\,\sigma_{ij;kl}\,\phi_k\,\chi_l
\ee
The following iteration scheme turns out in practice to be an
efficient numerical method.  It may not necessarily give a global
maximum, but at least it gives a useful local maximum that may depend
on a randomly chosen starting point.

The method is based on the observation that the maximum value of $s'$
is actually the maximal eigenvalue $\mu$ in the two linked eigenvalue
problems
\be{eq013}
\sum_k A_{ik}\,\phi_k=\mu\phi_i\;,\qquad
\sum_l B_{jl}\,\chi_l=\mu\chi_j
\ee
where
\be{eq014}
A_{ik}=\sum_{j,l}\chi_j^{\ast}\,\sigma_{ij;kl}\,\chi_l\;,\qquad
B_{jl}=\sum_{i,k}\phi_i^{\ast}\,\sigma_{ij;kl}\,\phi_k
\ee
Thus, we may start with any arbitrary unit vector
$|\chi\rangle=\sum_j\chi_j\,|j\rangle_B\in{\cal H}_B$ and compute the
hermitian matrix $A$.  We compute the unit vector
$|\phi\rangle=\sum_i\phi_i\,|i\rangle_A\in{\cal H}_A$ as an
eigenvector of $A$ with maximal eigenvalue, and we use it to compute
the hermitian matrix $B$.  Next, we compute a new unit vector
$|\chi\rangle$ as an eigenvector of $B$ with maximal eigenvalue, and
we iterate the whole procedure.

This iteration scheme is guaranteed to produce a non-decreasing
sequence of function values $s'$, which must
converge to a maximum value $\mu$.
This is at least a local maximum, and there corresponds to it at
least one product vector $|\phi\rangle\otimes|\chi\rangle$
and product density matrix
$\rho_p=(|\phi\rangle\langle\phi|)\otimes(|\chi\rangle\langle\chi|)$.

The above construction of $\rho_p$ implies, if $s>0$, that there exist
separable states
\be{eq015}
\rho'_s=(1-\lambda)\rho_s+\lambda\rho_p
\ee
with $0<\lambda\leq 1$, closer to $\rho$ than $\rho_s$ is.  However,
it turns out to be very inefficient to search only along the line
segment from $\rho_s$ to $\rho_p$ for a better approximation to
$\rho$.  It is much more efficient to append the new $\rho_p$ to a
list of product states $\rho_{pk}$ found in previous iterations, and
then minimize
\be{eq016}
F=\Tr\left(\rho-\sum_k\lambda_k\rho_{pk}\right)^2
\ee
which is a quadratic function of coefficients $\lambda_k\geq 0$ with
$\sum_k\lambda_k=1$.  We solve this quadratic programming problem by
an adaptation of the conjugate gradient method, and we throw away a
given product matrix $\rho_{pk}$ if and only if the corresponding
coefficient $\lambda_k$ becomes zero when $F$ is minimized.  In
practice, this means that we may construct altogether several hundred
or even several thousand product states, but only a limited number of
those, typically less than 100 in the cases we have studied, are
actually included in the final approximation $\rho_s$.

%%%%%%%%%%
\subsection{Bound entanglement in the $3\times 3$ system}
%%%%%%%%%%

For the $3\times 3$ system (composed of two three-level systems) there
are explicit examples of entangled states that remain positive under a
partial transposition. This was first discussed by Horodecki
\cite{Horodecki97} and then an extended set of states was found by
Bruss and Peres \cite{Bruss00}. We apply the method outlined above to
density matrices limited to a two dimensional planar section of the
full set. The section is chosen to contain one of the Horodecki states
and a Bell state in addition to the maximally mixed state, and the
method is used to identify the boundary of the separable states in
this two dimensional plane.

%%%%%%%%%%%%
\begin{figure}[h]
\begin{center}
\includegraphics[width=8cm]{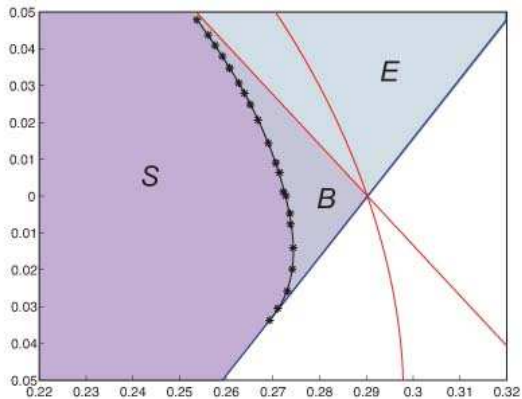}
\end{center}
\caption{{\small The boundary of the set of separable states in a two
    dimensional section, as determined by the numerical method. $\cS$
    denotes the set of separable states, $\cB$ the states with bound
    entanglement and $\cE$ the entangled states violating the Peres
    condition.  The blue straight line is determined by the algebraic
    equation $\det\rho=0$ and gives the boundary of the full set of
    density matrices. The red curves are determined by the equation
    $\det\rho_p=0$, in particular, the red straight line gives the
    boundary of the set of states that satisfy the Peres condition.
    The black dots are the numerically determined points on the
    boundary of the set of separable states. Note that the coordinates 
given in this plot differs from the distance defined in the main text by a 
factor $\sqrt{2}$.}}
\label{Bound}
\end{figure}
%%%%%%%%%%%%
Since the separable states $\cal S$ are contained in the 
set $\cP$ of
states that remain positive under partial transposition, we start the
search for the boundary of $\cS$ with a series of states located at
the boundary of $\cP$. This boundary is found by solving the algebraic
equations $\det\rho=0$ and $\det\rho^P=0$. For each chosen state we
find the distance to the closest separable state and change the test
state $\rho$ on a straight line between this point and the maximally
mixed state, in a step of length equal to the evaluated distance.  In
a small number of steps the intersection of the straight line and the
boundary of the separable states is found within a small error,
typically chosen to be $10^{-6}$. (The distance from the maximally
mixed state to the pure states in this case is $d=2/3$.)

In Fig.~\ref{Bound} we show a plot of the results of the calculations.
The numerically determined points on the border of the set of
separable states $\cS$ are indicated by black dots, while the border
of the set of states $\cP$ that satisfy Peres' condition, determined
by solving the algebraic equations, is shown as blue and red lines
which cross at the position of the Horodecki state. One should note
that the states with bound entanglement in the corner of the set $\cP$
cover a rather small area.

\section{Conclusions}

To summarize, we have in this paper focussed on some basic questions
concerning the geometry of separability. The simplest case of $2\times
2$ matrices has been used to demonstrate the usefulness of relaxing
the normalization requirement $\Tr\rho=1$. Thus, if this condition is
replaced by $\det\rho=1$, a relativistic description with a Minkowski
metric can be used, where all (non-pure) states can be connected by
Lorentz transformations. For a composite system consisting of two
two-level systems, independent Lorentz transformations performed on
the two sub-systems can be used to diagonalize an arbitrary density
matrix in a way that respects separability. We have used this
diagonalization to demonstrate the known fact that the Peres condition
and the separability condition are equivalent in this case.

Although the diagonalization with Lorentz transformations is
restricted to the composite $2\times 2$ system, we have shown that the
generalized form of the Schmidt decomposition used in this
diagonalization can be extended to higher dimensions. The
decomposition involves the use of non-unitary $SL(n,\mathbb C)$
transformations for the two subsystems. Although a full
diagonalization is not obtained in this way, we suggest that the
Schmidt decomposed form may be of interest in the study of
separability and bound entanglement.

A third part of the paper has been focussed on the use of a numerical
method to study separability. This method exploits the fact that the
set of separable states is convex, and is based on an iterative scheme
to find the closest separable state for an arbitrary density matrices.
We have demonstrated the use of this method in a numerical study of
bound entanglement in the case of a $3\times 3$ system. A further
study of separability with this method is under way.

\vspace{\baselineskip}

{\bf Acknowledgments}

We acknowledge the support of NorForsk under the 
Nordic network program 
{\em Low-dimensional physics: The theoretical 
basis of nanotechnology}. 
One of us (JML) would like to thank Prof. 
B.~Janko at the Institute for 
Theoretical Sciences, a joint 
institute of University of Notre Dame and the 
Argonne Nat.\ Lab., for 
the support and hospitality during a visit in the fall of 2005.

\end{document}